\documentclass[acmsmall,nonacm]{acmart}
\newcommand{\added}[1]{#1}

\newcommand{\revadd}[1]{#1}
\newenvironment{revisionadd}{}{}
\ifdefined\ACMMarkedRevision
  \newcommand{\rTwoOne}[1]{\textcolor{blue}{#1}}
  \newcommand{\rTwoThree}[1]{\textcolor{blue}{#1}}
\else
  \newcommand{\rTwoOne}[1]{#1}
  \newcommand{\rTwoThree}[1]{#1}
\fi
\usepackage{makecell}
\usepackage{algorithm}
\usepackage{algorithmic}

\AtBeginDocument{%
  }

\setcopyright{none}
\acmJournal{TODAES}
\acmVolume{}
\acmNumber{}
\acmArticle{}
\acmDOI{}

\begin{document}

\title[Formal-Feedback BO for LMC]{Formal-Verification-Guided Bayesian Optimization for Non-Intrusive Last-Mile Coverage Closure}
\thanks{This work is partially supported by \grantsponsor{NSF}{NSF}{https://www.nsf.gov/} (\grantnum{NSF}{CCF-2106725}, \grantnum{NSF}{CCF-2212346}, \grantnum{NSF}{CCF-2425399}, \grantnum{NSF}{CCF-2529764}), \grantsponsor{TAMU}{Texas A\&M University}{https://www.tamu.edu/} \grantnum{TAMU}{TPT Fund}, and \grantnum{TUMIAS}{Hans Fischer Senior Fellowship} by \grantsponsor{TUMIAS}{Technical University of Munich Institute for Advanced Study}{https://www.ias.tum.de/}.}

\author{Chengjia Liu}
\orcid{0009-0007-9711-5312}
\affiliation{%
  \department{Department of Electrical and Computer Engineering}
  \institution{Texas A\&M University}
  \city{College Station}
  \state{Texas}
  \country{USA}}
\email{chengjialiu@tamu.edu}

\author{Priyadarsi Mishra}
\orcid{0009-0002-6689-7036}
\affiliation{%
  \department{Department of Computer Science and Engineering}
  \institution{Texas A\&M University}
  \city{College Station}
  \state{Texas}
  \country{USA}}
\email{priyadarsimishra@tamu.edu}

\author{David Kebo Houngninou}
\orcid{0000-0002-7017-8440}
\affiliation{%
  \department{Department of Computer Science and Engineering}
  \institution{Texas A\&M University}
  \city{College Station}
  \state{Texas}
  \country{USA}}
\email{davidkebo@tamu.edu}

\author{Michael Quinn}
\orcid{0000-0002-6053-670X}
\affiliation{%
  \department{Department of Computer Science and Engineering}
  \institution{Texas A\&M University}
  \city{College Station}
  \state{Texas}
  \country{USA}}
\email{m.d.quinn@tamu.edu}

\author{Aakash Tyagi}
\orcid{0000-0002-9992-7628}
\affiliation{%
  \department{Department of Computer Science and Engineering}
  \institution{Texas A\&M University}
  \city{College Station}
  \state{Texas}
  \country{USA}}
\email{tyagi@tamu.edu}

\author{Jiang Hu}
\orcid{0000-0003-1157-7799}
\affiliation{%
  \department{Department of Electrical and Computer Engineering}
  \institution{Texas A\&M University}
  \city{College Station}
  \state{Texas}
  \country{USA}}
\email{jianghu@tamu.edu}

\renewcommand{\shortauthors}{Liu et al.}

\begin{abstract}
\begin{revisionadd}
Last-mile functional coverage closure for modern chip designs remains costly because a mature UVM environment exposes many controls while direct hits on the remaining bins are sparse. We present a non-intrusive workflow that closes \emph{pre-defined bins in an existing, human-authored functional coverage model}; it does not synthesize verification objectives from the design specification. The workflow first uses abstract-syntax-tree (AST) parsing and cone-of-influence (COI) tracing to map each residual target to existing testbench knobs. Under a matched 20-simulation budget, Bayesian optimization (BO) is compared with continued constrained-random verification (CRV) and random search over the same AST/COI-reduced space. Formal reachability then removes only targets supported by complete unreachability proofs, after which retained targets are grouped by formal difficulty and processed by group-specific BO under a second matched 20-simulation budget. Cleanup and grouping never add numerator bins; only simulator coverage artifacts do. On OpenTitan AES, \rTwoThree{first-pass reduced-space BO reaches 241/271 bins (88.93\% raw coverage); formal cleanup removes two denominator bins; the integrated post-cleanup stage---difficulty grouping, policy routing, and group-specific BO---adds 20 simulator-confirmed hits, yielding 261/271 raw coverage (96.31\%) and 261/269 effective coverage (97.03\%).} On Entropy Source, \rTwoThree{first-pass reduced-space BO reaches 2117/2277 bins (92.97\% raw coverage); formal cleanup removes ten denominator bins; the integrated post-cleanup stage---difficulty grouping, policy routing, and group-specific BO---adds 93 simulator-confirmed hits, yielding 2210/2277 raw coverage (97.06\%) and 2210/2267 effective coverage (97.49\%).} \rTwoThree{These results characterize the integrated diagnosis-and-search workflow; the present evaluation does not isolate the causal contributions of difficulty classification, policy routing, and post-cleanup BO.}
\end{revisionadd}
\rTwoThree{All comparisons are observations from single matched fixed-budget runs, not replicated statistical estimates.}
\end{abstract}



\maketitle

\section{Introduction}
\label{sec:intro}

\revadd{Functional verification remains a major cost and schedule driver in modern digital chip design, accounting for up to 70\% of project resources \cite{VerificationCost}. UVM-based constrained-random verification provides broad exploration \cite{UVM}, but sign-off often leaves a sparse set of hard-to-reach bins associated with corner cases, deep state interactions, and rare event sequences \cite{LMTChallenge}. These last-mile bins resist further undirected simulation and typically require expert analysis and directed tests, increasing closure time and the risk that coverage holes remain unresolved \cite{Wu2024SurveyHardwareDV}.}

\revadd{Existing automation only partly addresses this problem. Directed sequences and constraint overrides can be effective but require expert effort and may destabilize mature environments \cite{piziali2007functional}; coverage-feedback CRV may still require extensive simulation \cite{bergeron2006coverage}; and solver-based stimulus generation can require scenario-specific models \cite{ConstraintSolvers}. Formal methods provide exhaustive reasoning for focused properties but face full-chip scalability limits \cite{6418243,FormalScalability}. ML-assisted verification is promising \cite{FGG,Masamba2022SupervisedSelection,Nazi2022AdaptiveTestGeneration,gogri2020machine}, yet sparse target hits and expensive evaluations can limit fine-grained targeting and sample efficiency \cite{MLChallenges}.}

\begin{revisionadd}
To address these limitations, we formulate LMC closure as a staged diagnosis-and-search problem over an existing coverage plan. The optimization targets are the coverpoints and crosses already present in the verification environment. The method does not infer missing objectives from the specification, generate SVA assertions, or establish that the coverage plan itself is complete. Accordingly, our results quantify progress only on the supplied functional coverage model.

The end-to-end workflow is: (1) build a common initial coverage database from the unmodified CRV regressions; (2) extract the residual bins; (3) map each target to executable testbench controls with AST/COI analysis; (4) compare continued CRV, reduced-space random search, and reduced-space BO under the same 20-simulation budget; (5) remove only bins with complete, assumption-audited proofs of unreachability; (6) classify retained targets as constraint-heavy, temporal-heavy, or inconclusive; (7) run the corresponding group-specific BO policy; and (8) audit every numerator change against simulator coverage feedback. Formal cleanup changes only the effective denominator, and difficulty grouping changes only the search policy.

\paragraph{Relationship to the conference version.}
The preliminary MLCAD paper~\cite{liu2025lmc} contributed the non-intrusive AST/COI linkage front-end and a cGAN-based search stage. This journal article retains that structural front-end but replaces the cGAN back-end with target-conditioned BO, inserts complete-proof-only formal cleanup, adds formal-difficulty routing for the retained targets, and audits the resulting simulations at bin level. It also adds matched CRV and reduced-space-random baselines, raw/effective accounting, an assumption audit, resource measurements, objective-weight sensitivity experiments, and an end-to-end target case study. The remainder of this paper is self-contained and uses the conference work only to identify the retained component.

The main contributions are:
\begin{itemize}
  \item a precise coverage-model scope and accounting rule that separates simulator-confirmed hits from complete-proof denominator refinement;
  \item a reproducible target-conditioned BO configuration, including mixed-knob encoding, cold-start and ExtraTrees surrogates, difficulty-conditioned acquisition, objective terms, and stopping rules;
  \item a semi-automatic formal flow with mechanically generated cover properties, explicit environment assumptions, complete-proof-only cleanup, and constraint-heavy/temporal-heavy policy routing;
  \item matched 20-simulation baselines and sensitivity experiments on OpenTitan AES and Entropy Source, with final raw coverage of 261/271 and 2210/2277 bins, respectively; and
  \item quantitative COI funnels, formal/runtime statistics, and a worked AES bin from static linkage through simulator-confirmed closure.
\end{itemize}
\end{revisionadd}

\section{Background and Related Work}
\label{sec:background}

\subsection{Static Analysis for Verification Guidance}
\label{subsec:static_analysis}

Static analysis techniques, which examine code structure without executing simulations, offer valuable insights for guiding verification efforts. Abstract Syntax Trees (ASTs) provide a hierarchical representation of RTL and testbench source code, enabling automated parsing and structural understanding \cite{VeribleTool,4339230}. Cone of Influence (COI) tracing is another crucial technique that identifies the logical dependencies within a design, determining which signals or registers can influence (or are influenced by) a specific point of interest \cite{10958410}.

In the context of coverage closure, static analysis holds strong potential for bridging the gap between high-level testbench controls (knobs) and low-level Design Under Test (DUT) behavior related to specific coverage points. While manual analysis often involves implicitly tracing such dependencies, automating this through AST parsing and COI tracing can systematically identify candidate knobs relevant to specific coverage goals \cite{athavale2011evaluating}. For instance, COI can trace back from a coverage point's sampled signals to DUT inputs or configuration registers, and further analysis of the testbench AST can link these interfaces back to controlling knobs. This structural linkage information, although potentially complex to extract reliably for all scenarios, can significantly prune the search space for targeted stimulus generation compared to treating the entire knob configuration space as a black box. \revadd{COI is used only for dependency tracing and search-space reduction; it is not a coverage metric. Every coverage number reported in this paper comes from the DUT functional coverage model.}

\added{Recent survey work on directed test generation and ML-assisted hardware verification reinforces the importance of structure-aware guidance in reducing verification cost and improving search efficiency \cite{Jayasena2024DirectedTestSurvey,Wu2024SurveyHardwareDV}. In particular, Nazi et al. proposed an adaptive coverage-closure framework that automatically extracts a coverage dependency graph and uses this dependency information to bias stimulus generation toward faster functional coverage closure \cite{Nazi2022AdaptiveTestGeneration}. These results support the broader observation that explicit dependency information is especially useful when coverage holes are sparse and the controllable verification space is large.} 

\subsection{Machine Learning and Optimization for Stimulus Generation}
\label{subsec:ml_stimulus}

\revadd{Machine learning} (ML) has emerged as a promising avenue for automating aspects of stimulus generation, aiming to improve efficiency beyond traditional constrained-random methods \cite{Nazi2022AdaptiveTestGeneration,gogri2020machine}. Various approaches have been explored:

\begin{itemize}
    \item \textit{Reinforcement Learning (RL):} RL agents have been trained to learn policies for generating test sequences or configuring parameters to maximize coverage rewards \cite{10745410}. However, RL often requires a large number of simulation interactions and careful reward shaping, which can be challenging for sparse LMC targets.

    \item \textit{Genetic Algorithms (GA):} GAs are heuristic search algorithms inspired by natural evolution. They maintain a population of candidate solutions (e.g., knob configurations) and iteratively improve them through selection, crossover, and mutation operators based on a fitness function (e.g., achieved coverage or proximity to target) \cite{subedha2012efficient,jiawen2013coverage}. GAs are well-suited for exploring large search spaces and handling various knob types, but they typically require evaluating many candidate solutions per generation, impacting sample efficiency when simulations are costly.

    \item \textit{Generative Models:} Generative models aim to learn the underlying distribution of ``effective'' stimuli. Generative Adversarial Networks (GANs) \cite{gulrajani2017improvedtrainingwassersteingans} are particularly relevant due to their ability to generate complex, high-dimensional data. A GAN consists of a Generator network that creates synthetic data and a Discriminator network that distinguishes synthetic data from real data or, in adapted forms, evaluates its quality. Conditional GANs (cGANs) \cite{smith2020conditional} further allow the generation process to be guided by specific input conditions such as a target coverage bin. While GANs have shown success in image and text generation, their application to generating structured, functional stimuli for hardware verification, especially targeting specific \textit{uncovered} states using limited historical data, presents unique challenges related to training stability, evaluation metrics, and domain-specific guidance integration.

    \item \revadd{\textit{Black-box and Bayesian Optimization:} A complementary perspective treats verification parameter tuning as an expensive black-box optimization problem. Huang et al. showed that simple black-box optimization can improve coverage by directly tuning test parameters \cite{Huang2022BlackboxOptimization}, while Kumar et al. explored constrained-random verification optimization using ML and Bayesian estimation \cite{Kumar2023BayesianEstimation}. Together, these works frame verification tuning as a budgeted search problem rather than purely as a generation problem.}
\end{itemize}

\added{Recent work has diversified this design space along several orthogonal directions. Gal et al. used deep neural networks and smart sampling to accelerate coverage closure and, in a companion work, used clustering to identify large coverage holes, demonstrating that learned statistical structure can prioritize productive regions of the verification space \cite{Gal2020DNN,Gal2020CoverageHoles}. Gogri et al. later explored transaction-level stimulus optimization using ML predictors \cite{Gogri2022TransactionLevel}. In parallel, RL-based approaches have also been explored for direct stimuli generation, including actor-critic-based methods for IC design verification \cite{Tweehuysen2023ActorCriticRL}.}

\added{A complementary line of work focuses on selecting which tests to simulate rather than synthesizing entirely new stimuli. Novelty-Driven Verification showed that novelty-based ordering of generated tests can reduce the time required to close coverage \cite{Blackmore2021NoveltyDriven}. Subsequent work used supervised learning for coverage-directed test selection \cite{Masamba2022SupervisedSelection}, hybridized coverage-directed and novelty-driven selection \cite{Masamba2022HybridTesting}, used neural networks for novelty-based test selection \cite{Zheng2023NoveltyNN}, extended novelty detection to temporal patterns \cite{Zheng2024TemporalNovelty}, and used unsupervised learning from fast functional simulation to reduce expensive RTL simulation effort \cite{Liang2023UnsupervisedRTLSelection}. These works are highly relevant because they confirm that coverage feedback can be exploited effectively even when direct target hits are sparse.}

\revadd{More recently, LLM-based frameworks such as AutoBench and LLM4DV have explored automated testbench and stimulus generation for hardware verification \cite{Qiu2024AutoBench,Zhang2025LLM4DV}. These works target a broader setup and generation problem than the specific last-mile coverage problem considered here. Our method builds on the structure-guided knob linkage introduced in \cite{liu2025lmc} and couples that non-intrusive front-end with formal reachability diagnosis and a BO search engine suited to sparse, expensive verification feedback.}

\subsection{Formal Verification for Reachability Diagnosis}
\label{subsec:formal_reachability}

Formal methods have long been used to reason about corner-case behaviors that are difficult to expose with simulation alone {\cite{6418243, 3620237.3620314}}. In principle, formal reasoning can determine whether a target state or property is reachable under a set of assumptions, making it an attractive complement to simulation-based coverage closure. At the same time, formal methods also face well-known scalability challenges when applied indiscriminately to large designs or to unconstrained full-chip state spaces \cite{FormalScalability}. For this reason, formal analysis is often most effective when used selectively and in a focused manner. \rTwoOne{Recent learning-assisted formal-verification work has studied operator-level reasoning trees and rule learning as a complementary direction to formal analysis \cite{Wan2026FVRuleLearner}.}

\revadd{In this work, we do not use formal verification as a wholesale replacement for simulation-based coverage closure. Instead, we use it as a targeted diagnostic oracle inside the closure loop. Its first role is reachability triage: after the first-pass search and before second-stage simulation budget is spent on a residual case, the formal engine is asked whether the target is provably unreachable under the current design and environment assumptions. Its second role is difficulty diagnosis: for reachable hard cases, formal artifacts such as witness structure, depth, or enabling-condition patterns can provide clues about whether the dominant blocker is a narrow conjunction of value constraints or a long sequential dependency. This use of formal verification is important because coverage reports alone do not explain \emph{why} a bin remains uncovered.}

\revadd{The distinction between unreachable, constraint-heavy, and temporal-heavy cases is particularly valuable in late-stage closure. An unreachable case should not remain in the optimization queue. A reachable case dominated by a tight conjunction of predicates benefits from a different search policy than one dominated by a long protocol-sensitive event sequence. This triage addresses a limitation of simulation-only search by making that distinction explicit before the post-cleanup search.}

\subsection{Prior Work and Positioning}
\label{subsec:previous_works}

\revadd{Recent work relevant to our method can be grouped into four categories. First, \emph{structure- or dependency-guided stimulus optimization} uses design knowledge to bias search, as exemplified by coverage dependency graph extraction for coverage closure \cite{Nazi2022AdaptiveTestGeneration}. Second, \emph{learning-assisted coverage acceleration and hole prioritization} uses predictors, clustering, or novelty detection to characterize the coverage space or rank tests \cite{Gal2020DNN,Gal2020CoverageHoles,Blackmore2021NoveltyDriven,Zheng2023NoveltyNN,Zheng2024TemporalNovelty,Liang2023UnsupervisedRTLSelection}. Third, \emph{optimization-based tuning of verification parameters} formulates verification control as a black-box search problem and directly optimizes knobs or parameters \cite{Huang2022BlackboxOptimization,Kumar2023BayesianEstimation}. Fourth, \emph{direct policy learning and generative automation} uses RL or LLMs to generate effective stimuli or testbenches \cite{Tweehuysen2023ActorCriticRL,Qiu2024AutoBench,Zhang2025LLM4DV}.}

\revadd{Compared with these works, our method operates non-intrusively through existing high-level testbench knobs discovered by AST/COI analysis, rather than rewriting the testbench or treating the entire stimulus space as an undifferentiated black box. It addresses three limitations of simulation-only generative search in the true last-mile regime: (1) some bins remaining after first-pass search are unreachable and should be filtered before second-stage simulation effort is spent on them; (2) the remaining reachable hard bins are heterogeneous, with some dominated by value-conjunction difficulty and others by temporal sequencing difficulty; and (3) when each new evaluation is expensive and direct hits are sparse, sample-efficient black-box optimization can be a better fit than full generative modeling.}

\section{Methodology}
\label{sec:method}

\begin{figure*}[!t]
    \centering
    \includegraphics[width=\textwidth]{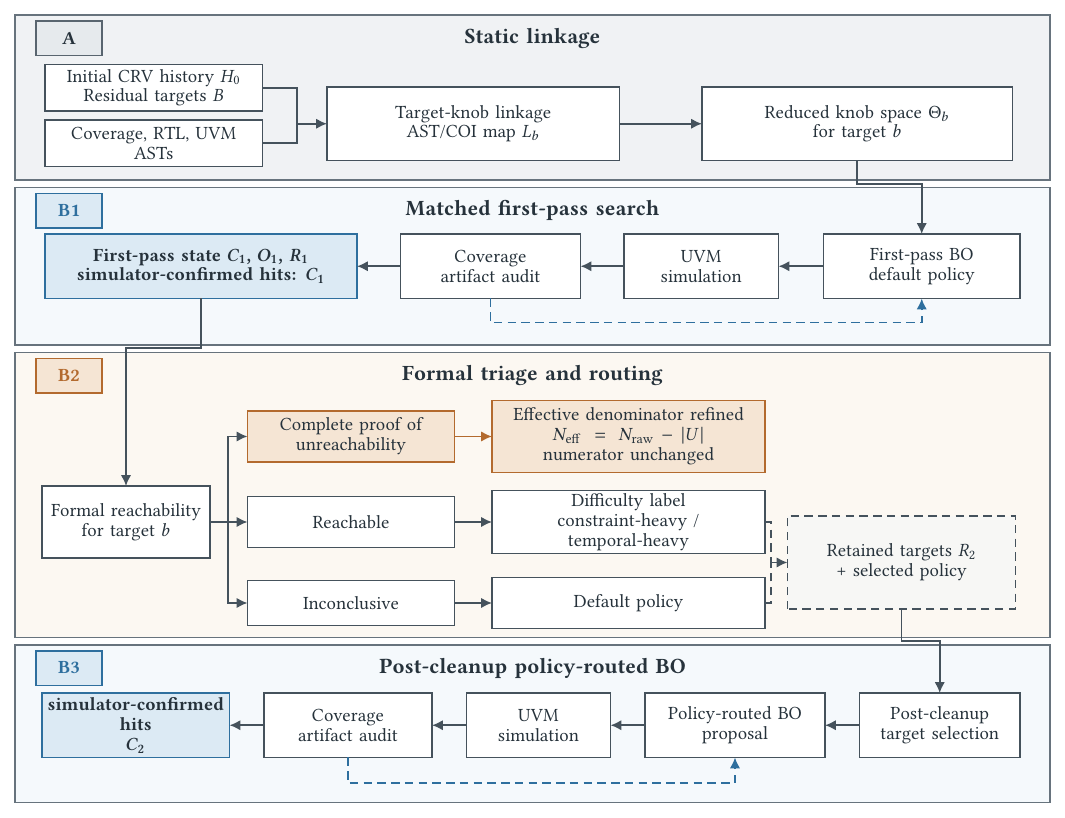}
    \caption{Overview of the proposed methodology. \rTwoOne{The workflow is organized into static linkage (A), matched first-pass search (B1), formal triage and routing (B2), and post-cleanup policy-routed BO (B3). The first-pass and post-cleanup loops report simulator-confirmed hits $C_1$ and $C_2$, respectively. Complete unreachability proofs refine the effective denominator without changing the numerator; reachable targets receive a difficulty label and selected policy before post-cleanup BO.}}
    \Description{A four-stage workflow for non-intrusive last-mile coverage closure. Stage A performs static target-knob linkage using coverage, RTL, and UVM ASTs. Stage B1 runs matched first-pass Bayesian optimization with UVM simulation and coverage-artifact audit. Stage B2 performs formal reachability triage, refines the effective denominator only for complete unreachability proofs, and assigns difficulty labels and policies to retained targets. Stage B3 runs post-cleanup policy-routed Bayesian optimization with simulator-confirmed coverage feedback.}
    \label{fig:overall_workflow}
\end{figure*}

Our proposed framework for accelerating last-mile functional coverage closure integrates static code analysis, \added{formal-verification feedback, and difficulty-aware Bayesian optimization}. The workflow, depicted in Figure~\ref{fig:overall_workflow}, consists of two primary stages: (A) Static Analysis for Testbench Knob to Coverage Linkage, and (B) \added{Formal-Feedback-Guided Knob Search with Reachability Triage and Bayesian Optimization}.

\subsection{Static Analysis for Testbench Knob to Coverage Linkage}
\label{subsec:static_analysis_linkage}

\revadd{This stage maps each uncovered target to a reduced set of high-level testbench knobs with a plausible structural influence on that target. The mapping is generated in four steps:}

\begin{enumerate}
    \item \revadd{\textbf{Target Coverage Bin Identification and DUT Signal Mapping:} Experts select residual targets from coverage databases produced by prior simulations. An automated SystemVerilog parser then locates each selected \texttt{coverpoint} or \texttt{cross} and extracts the DUT signals or internal state variables sampled by that definition. The output is the set of sampled DUT expressions used to seed backward analysis.}

    \item \revadd{\textbf{AST Parsing:} Verible \cite{VeribleTool} parses the DUT RTL and UVM testbench sources into ASTs. These structured representations support programmatic traversal of RTL dependencies and testbench control paths.}

    \item \revadd{\textbf{COI Tracing within the DUT:} Starting from the sampled DUT expressions, backward COI tracing identifies upstream signals, registers, and DUT ports that can affect the target. The retained \emph{Influential DUT Interfaces} comprise primary inputs, configuration registers, and other externally controllable points with a structural path to the target condition.}

    \item \revadd{\textbf{Testbench Linkage Analysis:} The testbench AST is traced from each influential DUT interface through drivers or Bus Functional Models (BFMs), sequences, configuration paths, and, where applicable, the Register Abstraction Layer (RAL). The analysis follows randomized fields and controls exposed through mechanisms such as \texttt{uvm\_config\_db} or direct parameterization, and returns the existing high-level knobs that can drive the retained interfaces.}
\end{enumerate}

\revadd{The output is a per-target linkage map of existing testbench knobs with plausible structural paths to the target condition. This map defines the reduced search space used downstream without introducing new testbench control logic, preserving non-intrusive use of mature verification environments.}

\begin{revisionadd}
For quantitative reporting, we distinguish three levels: \emph{target-linked knobs} retained by static linkage, \emph{executable knobs} actually admitted to a proposal, and the resulting \emph{encoded features}. One-hot expansion may make the feature dimension larger than the physical knob count; it is therefore reported separately rather than described as additional control.
\end{revisionadd}

\subsection{Formal Reachability and Difficulty Triage}
\label{subsec:formal_triage}
\begin{revisionadd}
For a coverage bin $b$, the coverage parser extracts the sampled DUT expressions and a predicate $\phi_b$. For a cross, $\phi_b$ is the conjunction of the participating component-bin predicates. The generator emits a cover property labeled \texttt{c\_lmc\_target}, sampled on the covergroup clock and disabled during reset; the audited AES configuration uses \texttt{@(posedge clk\_i) disable iff (!rst\_ni)}. A wrapper and a bind statement are generated when necessary so that the property observes the same DUT/testbench point as the original coverage model, and JasperGold is invoked through generated Tcl.

\rTwoThree{The parser supports ordinary value bins and simple crosses when their source expressions, predicates, sampled signals, and target identifiers can be retained. For a cross, the component-bin predicates are conjoined. An \texttt{iff} condition is included only when its clock, reset, disable condition, and formal signal mapping agree with the coverage model.}

\rTwoThree{The workflow excludes \texttt{ignore\_bins} before forming the reported residual inventories of 44 AES and 202 Entropy Source targets. It does not use \texttt{illegal\_bins} as automatic reachability targets; these cases require manual review because tool-specific illegal-event semantics are not equivalent to proven unreachability. Transition and wildcard bins are excluded from automatic translation and sent to manual review; their temporal or four-state mask semantics are not inferred from a single-cycle predicate. Parameterized or array-based bins did not occur in the evaluated target sets, so this evaluation makes no automatic-translation claim for those forms. Testbench-side sampled expressions occur in the evaluated coverage models. They are translated only when every identifier maps to formal-visible state under the same sampling and disable conditions; a case that failed this mapping check would be sent to manual review rather than translated automatically.}

The parser assigns three distinct outcome semantics. A reached cover trace is a constructive reachability witness and keeps the bin in the original denominator. A terminal complete proof that \texttt{c\_lmc\_target} cannot be reached is eligible for removal from the \emph{effective} denominator after assumption audit. Bounded-only results, timeouts, unknown results, and tool errors are inconclusive: they remain in the denominator and never count as hits. The implementation records these outcomes under the internal label \texttt{UnknownReachable}; this label denotes an inconclusive result and does not assert that the target is reachable. Witness depth and bounded proof depth are used only to diagnose targets already known to be reachable; they are never used as unreachability evidence.

Only targets with canonical reachable status enter the difficulty classifier. Let $d$ and $e$ denote witness depth and temporal-event count (with missing values mapped to zero), $M$ the temporal signal-name hit count, $P$ the predicate count computed from the original target condition, and $K_t$ and $K_c$ the numbers of target-linked knobs categorized as temporal and constraint, respectively. The implementation sets $P=0$ for an empty condition and otherwise sets $P$ to at least one based on the total number of occurrences of \texttt{\&\&}, \texttt{||}, \texttt{inside}, and \texttt{==}. $M$ counts distinct matches from the fixed signal-name lexicon \{\texttt{state}, \texttt{fsm}, \texttt{event}, \texttt{delay}, \texttt{timeout}, \texttt{wait}, \texttt{cycle}, \texttt{backpressure}, \texttt{stall}, \texttt{seq}\} across sampled, hierarchical, and COI-interface names. With thresholds $D_{th}=20$ and $E_{th}=3$, the scores are\begin{equation}\begin{aligned}s_T &= \mathbf{1}[d>D_{th}] + \mathbf{1}[e>E_{th}] + M + K_t, \\s_C &= P + K_c + \mathbf{1}[d\leq D_{th}].\end{aligned}\label{eq:triage_scores}\end{equation}A reachable target is labeled \textsc{TemporalHeavy} iff $s_T>s_C$; otherwise, including ties, it is \textsc{ConstraintHeavy}. Inconclusive targets bypass scoring and use the default search policy. Property generation, Tcl generation, log parsing, and initial labeling are scripted, while proof completeness and assumption adequacy are audit-reviewed; the stage is therefore semi-automatic.

All formal runs used Cadence JasperGold Jasper Apps 2024.09 FCS. Each target is formulated as a cover-property reachability query. A generated trace establishes reachability. Failure to find a trace within a bounded depth is treated as inconclusive; only a terminal complete proof of unreachability under the audited assumptions permits denominator removal. A finite witness or proof depth is therefore never used by itself as the cleanup criterion. The environment declares \texttt{clk\_i} as the default formal clock, models the active-low reset through \texttt{rst\_ni}, constrains TL-UL requests to protocol-legal transactions, and restricts CSR accesses to software-permitted sequences and legal write ordering. Data values within legal transactions, entropy/EDN values, alert responses, irrelevant lifecycle/OTP values, and outputs of explicitly black-boxed primitives remain unconstrained.

To audit over-constraint, we inspected generated Tcl, SVA assumption modules, run manifests, and elaboration logs; cataloged each \texttt{assume}/\texttt{restrict}; checked that no assumption directly fixed a target predicate or forced its antecedent false; and selectively relaxed the TL-UL legality and CSR-ordering assumption groups while comparing status, witnesses, depth, and vacuity. Reachable properties retained witnesses, all 12 complete proofs used for cleanup remained valid, and no complete proof was introduced solely by an environmental assumption.

\rTwoThree{For each generated property, the source coverage target and generated formal target are compared through the original bin predicate, the generated SVA condition, the sampling clock, the reset/\texttt{disable iff}/\texttt{iff} conditions, and the signal mapping and bind connection. The proof status, witness or proof depth, vacuity result, and assumption audit are retained with the same target. Bounded, timed-out, unknown, and tool-error outcomes remain inconclusive and are not used for denominator removal.}

\rTwoThree{The reported accounting follows the same target set through each stage. The initial residual inventory contains 44 AES and 202 Entropy Source targets (Table~\ref{tab:initial_base}); first-pass BO covers 14 and 42 unique bins, leaving 30 and 160 targets for formal queries (Table~\ref{tab:matched_search} and Table~\ref{tab:cost_scale}). In these evaluated post-first-pass residual sets, no target required manual-review exclusion: all 30 AES and 160 Entropy Source targets were within the supported subset, were translated into formal queries, and were run. The supported-subset restrictions above therefore do not reduce the reported query counts, and no target was silently dropped by the parser. The formal outcomes are 22 reachable, 2 complete-unreachability, and 6 inconclusive for AES, and 94, 10, and 56, respectively, for Entropy Source. Thus, removing only the complete proofs leaves 28 and 150 retained targets.}
\end{revisionadd}

\subsection{Formal-Feedback-Guided, Difficulty-Aware Bayesian Optimization}
\label{subsec:bo_search}
\begin{revisionadd}
The BO state consists of \texttt{SimulationObservation} records containing the complete knob assignment, seed, target identifier, simulator validity/status, target hit, stepping/proximity measurements, formal alignment, simulation cost, coverage metadata, and objective value. Coverage-history summaries derived from the 3,000-run AES and 10,000-run Entropy Source CRV regressions provide the warm-start source. The corpus sizes are not surrogate-training observation counts. Candidate-specific observations are accumulated online; exact-target observations are preferred, and only valid, policy-credit-eligible observations with the same target-constraint signature or target family may be transferred.

Categorical knobs and explicit choices are one-hot encoded; Boolean knobs without choices use one bit; integer, ordinal, and floating-point knobs are min--max normalized. Candidate generation remains discrete: small integer ranges are enumerated, while integer ranges wider than eight and floating-point ranges use their minimum, midpoint, and maximum. Missing values use the configured default, and compatibility metadata plus target constraints remove invalid combinations before acquisition scoring.

With fewer than three eligible matching observations, or with a constant observed objective, the surrogate is \texttt{distance\_idw\_v1}. It selects the five nearest samples, uses $w_i=1/(d_i+10^{-6})$, attenuates the weighted objective mean by $1-0.5\min(d_{\min},1)$, and estimates uncertainty as $\min(1,d_{\min}+\max_i y_i-\min_i y_i)$. Otherwise, the implementation fits \texttt{ExtraTreesRegressor} with 64 trees, \texttt{min\_samples\_leaf=1}, and a deterministic target-ID-derived seed. Its uncertainty is the estimator prediction range plus nearest-sample distance, clipped to one. The surrogate is rebuilt before every proposal.

The default acquisition is difficulty-conditioned UCB:
\begin{equation}
a(\mathbf{x})=\mu(\mathbf{x})+\beta_{\tau}u(\mathbf{x})+0.25N(\mathbf{x})+0.10D(\mathbf{x})
+B_{\mathrm{diff}}(\mathbf{x})+B_{\mathrm{adj}}(\mathbf{x})+\epsilon(\mathbf{x}),
\label{eq:acquisition}
\end{equation}
where $N\in\{0,1\}$ is novelty, $D$ is batch diversity, and deterministic jitter satisfies $\epsilon\in[0,0.001)$. Constraint-heavy and inconclusive targets use $\beta_{\tau}=0.35$; temporal-heavy targets use $\beta_{\tau}=1.5$. The difficulty-conditioned term is $B_{\mathrm{diff}}(\mathbf{x})=\min(0.3,0.08n_t(\mathbf{x}))$ for a TemporalHeavy target and zero for ConstraintHeavy and inconclusive targets, where $n_t(\mathbf{x})$ is the number of active temporal knobs. In the evaluated configuration, $B_{\mathrm{adj}}(\mathbf{x})=B_{\mathrm{cs}}(\mathbf{x})+B_{\mathrm{hint}}(\mathbf{x})-p_{\mathrm{invalid}}(\mathbf{x})$. Here, $B_{\mathrm{cs}}\in[0,0.1]$ is the reward for satisfying active target or witness constraints, $B_{\mathrm{hint}}\in\{0,0.05\}$ marks condition- or witness-derived candidates, and $p_{\mathrm{invalid}}\in[0,1]$ is the learned invalid-execution probability. Candidates that violate hard compatibility or active constraints are removed before scoring rather than penalized through $B_{\mathrm{adj}}$. The configuration uses a per-restart acquisition-sampling cap of 1,024 candidates, four restarts, and retains the top eight candidates. This cap is not the size of the raw combinational pool.

After simulation, the objective is
\begin{equation}
J(\mathbf{x},b)=1.0H+0.3S+0.2F-0.001C-I_{\mathrm{invalid}},
\label{eq:objective}
\end{equation}
where $H=1$ only when simulator coverage feedback records the target hit; $S$ is the arithmetic mean of available stepping measurements, including partial-predicate satisfaction, COI-signal toggling, coverage-export matching, and target raw count; and $C$ is simulation/feedback wall-clock time in seconds. $F=0$ when formal analysis does not establish reachability, $F=0.5$ when reachability is established without witness depth, and otherwise $F=1/(1+d_w/20)$. The invalid-execution penalty is one. The temporal-knob acquisition bonus is not part of $F$ or $J$.

\rTwoThree{The stepping components have the following implementation ranges: partial-predicate satisfaction is in $[0,1]$; COI-signal toggling is an indicator in $\{0,1\}$; coverage-export matching is a binary indicator when a valid matching row is present; and target raw count is a nonnegative integer copied from the coverage export. The objective averages the fields present in each observation rather than imposing a fixed four-component vector. The parser applies no clipping, logarithmic transform, or min--max normalization to target raw count; therefore $S$ is not intrinsically bounded by one when that component is present, and $\lambda_S=0.3$ does not cap its magnitude. In the AES case study, $S=(0+1+0+2)/4=0.75$, where the final $2$ is a raw count.}

\rTwoThree{In the evaluated BO configuration, target utility rewarded earlier prescribed sequence positions, reachable or inconclusive formal status, larger linked-knob sets, untried targets, positive same-family feedback, a clipped score from a one-candidate acquisition preview, and bounded profile-feedback and small-replay adjustments; it penalized repeated target observations, valid target- or family-level misses, tool errors, and missing linkage. The preview was enabled for BO proposals and bypassed only during fixed-knob or non-BO execution; sequence- and history-dependent terms contributed only when their corresponding inputs or observations were available. The active target with the largest utility is selected, with sequence rank and target identifier used as deterministic tie-breakers. The workflow then constructs and ranks candidate configurations for that target with its target-specific surrogate and acquisition rule. It does not perform one global ranking of all $(b,\mathbf{x})$ pairs. Except for this clipped preview score, raw acquisition values are interpreted only within the selected target's proposal step: the target-specific surrogate is rebuilt for each proposal, and no shared cross-target calibration or per-target normalization of the raw acquisition score is assumed. Covered targets are removed from the active set, while an optional target sequence can impose a prescribed order. The policy therefore has no formal round-robin, priority-aging, or starvation-prevention guarantee. Formal-difficulty labels are assigned before post-cleanup proposals, and the policy associated with the selected target is then used. The group-scheduling mode in the reported 20-simulation post-cleanup run is jointly interleaved.}

A target leaves the active search queue after a simulator-confirmed hit or an audited complete unreachability proof; a search stage stops when its fixed per-DUT simulation budget is exhausted. A reachable, bounded, timed-out, unknown, or tool-error outcome does not justify denominator removal. Constraint-heavy targets are routed to low-beta constraint-focused BO, temporal-heavy targets to high-beta temporal-aware BO, and inconclusive targets to the default policy.
\end{revisionadd}

\subsection{End-to-End Execution and Coverage Accounting}
\label{subsec:closed_loop}
\begin{revisionadd}
The execution order is fixed: initial CRV history $\rightarrow$ residual extraction $\rightarrow$ AST/COI reduction $\rightarrow$ matched first-pass search $\rightarrow$ complete-proof-only cleanup $\rightarrow$ residual difficulty grouping $\rightarrow$ group-specific BO $\rightarrow$ simulator-artifact audit. The first-pass BO result is the pre-cleanup numerator baseline. Cleanup may reduce only the effective denominator; grouping only selects a policy. A bin enters the numerator only when the simulator coverage export records it as hit.

Each proposed knob assignment is translated into an executable test without rewriting the UVM environment. Completed simulations update the observation history, and already covered targets are skipped in subsequent selection. This preserves the existing non-intrusive control surface while making the search and accounting stages explicit. Algorithm~\ref{alg:last_mile_closure} states the executable control flow. In the first-pass search, no target-specific formal result is yet available, so the formal-alignment term in Eq.~\eqref{eq:objective} is set to $F=0$ as defined above; the group-specific search uses the audited formal result.
\end{revisionadd}

\begin{algorithm}[!t]\caption{Audited post-regression last-mile closure workflow}\label{alg:last_mile_closure}\small\begin{algorithmic}[1]
\REQUIRE CRV history $\mathcal{H}_0$ and per-DUT stage budgets $T_1,T_2$
\STATE $\mathcal{B}\gets\textsc{ExtractResidual}(\mathcal{H}_0)$
\FORALL{$b\in\mathcal{B}$}  
\STATE $\mathcal{L}_b\gets\textsc{StaticLinkage}(b)$ 
\STATE $\Theta_b\gets\textsc{BuildSearchSpace}(\mathcal{L}_b)$
\ENDFOR
\STATE $(\mathcal{C}_1,\mathcal{O}_1)\gets\textsc{Search}(\mathcal{B},\{\Theta_b\},\textsc{Default},T_1,\mathcal{H}_0)$
\STATE $\mathcal{R}_1\gets\mathcal{B}\setminus\mathcal{C}_1$; $\mathcal{U}\gets\varnothing$
\FORALL{$b\in\mathcal{R}_1$}  
\STATE $q_b\gets\textsc{GenerateCover}(b,\mathcal{L}_b)$; $r_b\gets\textsc{RunAndAuditFormal}(q_b)$    
\IF{$r_b$ is an audited complete unreachability proof}    
\STATE $\mathcal{U}\gets\mathcal{U}\cup\{b\}$  
\ELSIF{$r_b$ has canonical reachable status}    
\STATE $g_b\gets\textsc{TemporalHeavy}$ if $s_T>s_C$, else $\textsc{ConstraintHeavy}$ \COMMENT{Eq.~\eqref{eq:triage_scores}}    
\STATE $p_b\gets\textsc{GroupPolicy}(g_b)$  
\ELSE    
\STATE $g_b\gets\textsc{Inconclusive}$; $p_b\gets\textsc{Default}$  
\ENDIF
\ENDFOR
\STATE $\mathcal{R}_2\gets\mathcal{R}_1\setminus\mathcal{U}$
\STATE $(\mathcal{C}_2,\mathcal{O}_2)\gets\textsc{Search}(\mathcal{R}_2,\{\Theta_b\},\{p_b\},T_2,\mathcal{H}_0\cup\mathcal{O}_1)$
\RETURN $\mathcal{C}=\mathcal{C}_1\cup\mathcal{C}_2$ and $\mathcal{U}$; report Eqs.~\eqref{eq:raw_coverage}--\eqref{eq:effective_coverage}
\STATE \textbf{function} \textsc{Search}($\mathcal{T},\{\Theta_b\},p,n,\mathcal{O}$)  
\STATE $\mathcal{C}'\gets\varnothing$  
\FOR{$i=1,\ldots,n$}    
\STATE $\mathcal{A}\gets\mathcal{T}\setminus\mathcal{C}'$    
\IF{$\mathcal{A}=\varnothing$} 
\STATE \textbf{break} 
\ENDIF    
\STATE $(b_i,\mathbf{x}_i)\gets\textsc{Propose}(\mathcal{A},\{\Theta_b\}_{b\in\mathcal{A}},p,\mathcal{O})$ \COMMENT{\rTwoThree{Target utility selects $b_i$ first; target-specific acquisition then ranks $\mathbf{x}_i$ (Sec.~\ref{subsec:bo_search}).}}    
\STATE $z_i\gets\textsc{Simulate}(\mathbf{x}_i)$    
\STATE $\mathcal{D}_i\gets\textsc{AuditCoverageExport}(z_i)\cap\mathcal{T}$    
\STATE $\mathcal{C}'\gets\mathcal{C}'\cup\mathcal{D}_i$    
\STATE $o_i\gets\textsc{BuildObservation}(\mathbf{x}_i,b_i,z_i,\mathcal{D}_i)$; evaluate Eq.~\eqref{eq:objective}    
\STATE $\mathcal{O}\gets\mathcal{O}\cup\{o_i\}$  
\ENDFOR  
\RETURN $(\mathcal{C}',\mathcal{O})$
\STATE \textbf{end function}
\end{algorithmic}
\end{algorithm}

\subsection{Reproducibility and Implementation Boundary}
\label{subsec:reproducibility}
\begin{revisionadd}
We do not release the full production implementation of the AST/COI and BO workflow because it relies on local integration scripts and licensed installations of Cadence Xcelium, IMC, and JasperGold. The evaluated DUTs and UVM environments are open-source OpenTitan blocks. The manuscript provides workflow pseudocode, property-generation rules, data schemas, exact BO parameters, candidate-pool semantics, objective definitions, formal outcome rules, configuration and accounting tables, and an end-to-end case study. This information supports independent reimplementation and evaluation, but not exact artifact-level reproduction or direct execution of our internal production workflow.
\end{revisionadd}

\section{Experimental Setup}
\label{sec:exp}
\begin{revisionadd}
The evaluation follows the workflow order in Section~\ref{subsec:closed_loop}. All search-only branches begin from the same precomputed coverage database and use the same 20-simulation budget per DUT. The sequential workflow then applies complete-proof-only cleanup to the first-pass BO state, groups every retained target, and runs group-specific BO under a separate per-stage budget. These targeted budgets apply only to last-mile bins left unresolved by the preceding conventional regression and are not presented as a run-for-run extrapolation of conventional CRV. The search-only branches are matched by simulation count rather than wall-clock time, and the complete-workflow endpoint is not presented as a 20-simulation matched baseline. This separation prevents a denominator change or a classification label from being reported as a new coverage hit.
\end{revisionadd}

\subsection{Benchmarks and Verification Environment}
\label{subsec:exp_setup}
\begin{revisionadd}
We evaluate the OpenTitan AES Core and Entropy Source IPs in their unmodified UVM environments~\cite{OpenTitanAESSpec,OpenTitanEntropySpec}. The initial coverage database is produced by the standard CRV regressions, before any targeted search. Table~\ref{tab:initial_base} defines the starting point and the LMC target sets. All branches reuse this database; the source regressions contain 3,000 and 10,000 runs, and these are not surrogate-observation counts.

\begin{table*}[!t]
\centering
\caption{Common initial coverage base. Residual bins define the LMC target set.}
\label{tab:initial_base}
\small
\begin{tabular}{>{\raggedright\arraybackslash}p{\dimexpr0.14\linewidth-2\tabcolsep\relax} >{\raggedright\arraybackslash}p{\dimexpr0.27\linewidth-2\tabcolsep\relax} >{\raggedleft\arraybackslash}p{\dimexpr0.12\linewidth-2\tabcolsep\relax} >{\raggedleft\arraybackslash}p{\dimexpr0.22\linewidth-2\tabcolsep\relax} >{\raggedleft\arraybackslash}p{\dimexpr0.1\linewidth-2\tabcolsep\relax} >{\raggedleft\arraybackslash}p{\dimexpr0.15\linewidth-2\tabcolsep\relax}}
\toprule
\textbf{DUT} & \textbf{Initial regression source} & \textbf{CRV runs} & \textbf{Covered} & \textbf{Total bins} & \textbf{Residual targets} \\
\midrule
AES & Unmodified OpenTitan UVM regression & 3,000 & 227 (83.76\%) & 271 & 44 \\
Entropy Source & Unmodified OpenTitan UVM regression & 10,000 & 2075 (91.13\%) & 2277 & 202 \\
\bottomrule
\end{tabular}
\end{table*}

Cadence Xcelium \texttt{xrun} 24.03-s013 executes the simulations, and IMC provides the coverage artifacts~\cite{CadenceXcelium,CadenceIMC}. Verible parses RTL and testbench SystemVerilog into ASTs~\cite{VeribleTool}; local scripts perform COI tracing, linkage, property/Tcl generation, log parsing, and BO orchestration.

Table~\ref{tab:cost_scale} reports benchmark size and the measurements available from the audited runs. Formal triage is applied after first-pass BO to the 30 remaining AES and 160 remaining Entropy Source targets. The resulting checks are target-specific IP/block-level queries, not full-chip exhaustive proofs. The AST/COI timings cover the complete residual target sets (44 AES and 202 Entropy Source targets); the AST-front-end values are approximate. \rTwoThree{Total BO-stage wall-clock, including simulation execution, is reported for} the 20-candidate BO stage, with mean simulation runtime listed separately for context.

\begin{table*}[!t]
\centering
\caption{Audited benchmark and measured stage costs. AST-front-end times are approximate; AST/COI times cover the full initial residual target set, formal and model-fit rows report per-query or per-candidate summary statistics, and the total BO-stage row covers the full 20-candidate stage. \rTwoThree{The total BO-stage wall-clock values include simulation execution and are not full end-to-end workflow runtimes.}}
\label{tab:cost_scale}
\footnotesize
\setlength{\tabcolsep}{5pt}
\renewcommand{\arraystretch}{1.08}
\begin{tabular}{>{\raggedright\arraybackslash}p{\dimexpr0.34\linewidth-2\tabcolsep\relax} >{\centering\arraybackslash}p{\dimexpr0.33\linewidth-2\tabcolsep\relax} >{\centering\arraybackslash}p{\dimexpr0.33\linewidth-2\tabcolsep\relax}}
\toprule
\textbf{Category} & \textbf{AES} & \textbf{Entropy Source} \\
\midrule
RTL files / RTL LOC & 37 / 14,739 & 18 / 10,511 \\
Testbench files / TB LOC & 59 / 7,597 & 37 / 9,196 \\
Total functional coverage bins & 271 & 2,277 \\
Residual target bins at start & 44 & 202 \\
Total environment knobs & 44 & 63 \\
Verible dependency edges & 515 & 1,653 \\
Approximate AST front-end time & 6.85 s & 3.28 s \\
COI/linkage back-end time & 0.50 s & 101.32 s \\
Total AST/COI wall-clock & 7.35 s & 104.60 s \\
Formal queries & 30 & 160 \\
Reachable formal outcomes & 22 & 94 \\
Complete unreachability proofs & 2 & 10 \\
Inconclusive formal outcomes & 6 & 56 \\
Median / maximum formal runtime & 11.25 / 14.94 s & 19.25 / 50.34 s \\
Median / maximum proof depth & 3 / 5 & 3 / 5 \\
Peak formal memory & 0.651 GB & 1.798 GB \\
Median BO model-fit time & 78.32 s & 132.82 s \\
Mean simulation runtime & 147.56 s/candidate & 718.93 s/candidate \\
Total 20-candidate BO-stage wall-clock & 1 h 18 min 03 s & 4 h 35 min 24 s \\
\bottomrule
\end{tabular}
\end{table*}

Runs used the Texas A\&M University Olympus cluster on a dual-socket AMD EPYC 7443 node (48 physical cores) running Red Hat Enterprise Linux 8.10. Slurm allocated four CPU cores; at most eight simulation jobs ran concurrently, whereas formal runs were serial.
\end{revisionadd}

\subsection{Workflow Variants Under Comparison}
\label{subsec:workflow_variants}
\begin{revisionadd}
The evaluation distinguishes standalone search baselines from the sequential closure workflow:
\begin{enumerate}
  \item \textbf{Continued CRV:} twenty fresh simulations using the existing UVM constraints and seed schedule, without target guidance.
  \item \textbf{AST/COI-reduced random search:} twenty simulations sampled randomly from the target-linked executable knob space.
  \item \textbf{AST/COI-reduced first-pass BO:} twenty simulations over the same reduced space, using the surrogate and acquisition in Section~\ref{subsec:bo_search}. This is the pre-cleanup numerator baseline.
  \item \textbf{Complete-proof cleanup:} only complete, audited unreachability proofs are removed from the effective denominator. This stage adds zero hits.
  \item \textbf{Residual grouping and group-specific BO:} the 28 retained AES targets and 150 retained Entropy Source targets are routed to their low-beta, high-beta, or default policies according to the formal outcome. A matched 20-simulation run per DUT is then executed; grouping itself adds zero coverage, and only simulator-confirmed feedback changes the numerator.
\end{enumerate}
The cGAN workflow in \cite{liu2025lmc} is not used as the primary matched experimental control; the controlled comparisons above share the DUT, starting database, legal stimulus space, and simulation budget.
\end{revisionadd}

\subsection{Evaluation Protocol}
\label{subsec:eval_protocol}
\begin{revisionadd}
Raw functional coverage is always reported over the fixed original denominator:\begin{equation}
\mathrm{FC}_{\mathrm{raw}}=\frac{N_{\mathrm{sim-hit}}}{N_{\mathrm{all\ bins}}}.
\label{eq:raw_coverage}\end{equation}Effective coverage is reported separately and removes only complete unreachability proofs:\begin{equation}
\mathrm{FC}_{\mathrm{effective}}=
\frac{N_{\mathrm{sim-hit}}}{N_{\mathrm{all\ bins}}-N_{\mathrm{complete\ proofs}}}.
\label{eq:effective_coverage}\end{equation}A reached cover trace, difficulty label, bounded result, timeout, unknown result, or manual review never changes either numerator. Bounded, timed-out, and otherwise inconclusive outcomes stay in the denominator.

The matched search branches use the same initial database, target constraints, stopping rules, seed schedule, candidate-generation schedule, and 20-simulation budget. They are matched by simulation count, not by wall-clock time. Hit counts denote unique newly covered target bins across the stated budget; because one simulation may cover multiple bins, the hit count may exceed the number of simulations. \rTwoThree{A direct hit is a newly covered residual bin whose target was selected for that proposal; a collateral (side) hit is another newly covered residual bin reported by the same simulation. Counts are deduplicated by each target's first newly covered observation, so the total includes both categories and is not the number of successful target-specific proposals.} The objective sensitivity runs likewise use the same retained targets, history, constraints, stopping rules, seeds, and candidate schedule; only the four objective weights change. Each setting is one fixed-budget run rather than a replicated statistical experiment.
\end{revisionadd}

\section{Results and Analysis}
\label{sec:results}
\begin{revisionadd}
This section reports matched search-only baselines first, then the sequential effects of formal cleanup, residual grouping, and post-cleanup BO. This order keeps raw numerator gains separate from effective-denominator refinement.

\subsection{Matched Search-Only Baselines}
\label{subsec:matched_baselines}
Table~\ref{tab:matched_search} compares the three search-only branches from the common initial state in Table~\ref{tab:initial_base}. Every covered-bin count in this table is backed by simulator coverage output.

\begin{table*}[!t]
\centering
\caption{Matched search-only baselines under the same initial database and 20-simulation budget. New hits are unique target bins; one simulation may cover multiple bins.}
\label{tab:matched_search}
\small
\begin{tabular}{>{\raggedright\arraybackslash}p{\dimexpr0.14\linewidth-2\tabcolsep\relax} >{\raggedright\arraybackslash}p{\dimexpr0.27\linewidth-2\tabcolsep\relax} >{\raggedleft\arraybackslash}p{\dimexpr0.18\linewidth-2\tabcolsep\relax} >{\raggedleft\arraybackslash}p{\dimexpr0.12\linewidth-2\tabcolsep\relax} >{\raggedleft\arraybackslash}p{\dimexpr0.29\linewidth-2\tabcolsep\relax}}
\toprule
\textbf{DUT} & \textbf{20-simulation branch} & \textbf{Initial covered} & \textbf{New hits} & \textbf{Final raw coverage} \\
\midrule
AES & Continued CRV & 227 & 1 & 228/271 (84.13\%) \\
AES & Random over AST/COI-reduced space & 227 & 3 & 230/271 (84.87\%) \\
AES & BO over AST/COI-reduced space & 227 & 14 & 241/271 (88.93\%) \\
Entropy Source & Continued CRV & 2075 & 13 & 2088/2277 (91.70\%) \\
Entropy Source & Random over AST/COI-reduced space & 2075 & 21 & 2096/2277 (92.05\%) \\
Entropy Source & BO over AST/COI-reduced space & 2075 & 42 & 2117/2277 (92.97\%) \\
\bottomrule
\end{tabular}
\end{table*}

\rTwoThree{Relative to random search over the same reduced space, first-pass BO adds 11 AES hits and 21 Entropy Source hits, equivalent to 4.06 and 0.92 percentage points of raw coverage. The larger AES difference and smaller Entropy Source difference are consistent with DUT-dependent behavior in these audited fixed-budget runs. A BO run over the full unreduced knob space was not available in the audited evidence and is not claimed.}

\subsection{Sequential Raw and Effective Accounting}
\label{subsec:sequential_accounting}
Six AES targets and 56 Entropy Source targets produced bounded, timed-out, or otherwise inconclusive formal outcomes and remain in the denominator. Table~\ref{tab:sequential_accounting} starts from the first-pass BO state. Formal cleanup adds no bins: it removes two AES and ten Entropy Source bins only after complete proofs, producing effective denominators of 269 and 2267.

\begin{table*}[!t]
\centering
\caption{Sequential accounting\rTwoThree{: only simulator-confirmed first-pass or post-cleanup BO executions add numerator bins, complete-proof cleanup changes only the effective denominator, and difficulty grouping changes only policy selection.}}
\label{tab:sequential_accounting}
\footnotesize
\setlength{\tabcolsep}{4pt}
\begin{tabular}{>{\raggedright\arraybackslash}p{\dimexpr0.115\linewidth-2\tabcolsep\relax} >{\raggedright\arraybackslash}p{\dimexpr0.22\linewidth-2\tabcolsep\relax} >{\raggedleft\arraybackslash}p{\dimexpr0.12\linewidth-2\tabcolsep\relax} >{\raggedleft\arraybackslash}p{\dimexpr0.12\linewidth-2\tabcolsep\relax} >{\raggedleft\arraybackslash}p{\dimexpr0.135\linewidth-2\tabcolsep\relax} >{\centering\arraybackslash}p{\dimexpr0.145\linewidth-2\tabcolsep\relax} >{\centering\arraybackslash}p{\dimexpr0.145\linewidth-2\tabcolsep\relax}}
\toprule
\textbf{DUT} & \textbf{Sequential stage} & \textbf{New simulator hits} & \textbf{Complete proofs} & \textbf{Inconclusive outcomes} & \textbf{Raw coverage} & \textbf{Effective coverage} \\
\midrule
AES & First-pass AST/COI + BO & 14 & -- & -- & 241/271 (88.93\%) & -- \\
AES & Complete-proof cleanup & 0 & 2 & 6 & 241/271 (88.93\%) & 241/269 (89.59\%) \\
AES & Post-cleanup group-specific BO & 20 & 2 & 6 & 261/271 (96.31\%) & 261/269 (97.03\%) \\
Entropy Source & First-pass AST/COI + BO & 42 & -- & -- & 2117/2277 (92.97\%) & -- \\
Entropy Source & Complete-proof cleanup & 0 & 10 & 56 & 2117/2277 (92.97\%) & 2117/2267 (93.38\%) \\
Entropy Source & Post-cleanup group-specific BO & 93 & 10 & 56 & 2210/2277 (97.06\%) & 2210/2267 (97.49\%) \\
\bottomrule
\end{tabular}
\end{table*}

\rTwoThree{After cleanup, all 28 retained AES targets and 150 retained Entropy Source targets are grouped and routed to the corresponding BO policy. The ensuing 20-simulation targeted stage produces 20 and 93 simulator-confirmed hits, respectively, yielding the final raw and effective values in Table~\ref{tab:sequential_accounting}. These hit counts are joint outcomes of retained-target grouping, group-policy selection, BO proposal, and simulation. A matched \rTwoOne{default-BO control after the same formal cleanup was not performed}; therefore, these aggregate gains do not isolate the contribution of difficulty classification or policy routing from the subsequent BO and simulation stage. Classification and cleanup by themselves add no numerator bins. In the audited post-cleanup runs, the 20 AES newly covered bins comprised 3 direct and 17 collateral hits, while the 93 Entropy Source newly covered bins comprised 17 direct and 76 collateral hits. The collateral hits are the expected consequence of stress-style sequences exercising many bins per run.}

\subsection{Quantitative AST/COI Reduction}
\label{subsec:coi_quant}
Table~\ref{tab:coi_funnel} reports representative target-specific AST/COI linkage funnels using a consistent count-based format for AES-T1 and the two Entropy Source targets.

\begin{table*}[!t]
\centering
\caption{Representative target-specific AST/COI linkage funnels. COI is a reduction mechanism, not a coverage metric.}
\label{tab:coi_funnel}
\footnotesize
\setlength{\tabcolsep}{3.5pt}
\begin{tabular}{>{\raggedright\arraybackslash}p{\dimexpr0.21\linewidth-2\tabcolsep\relax} >{\raggedright\arraybackslash}p{\dimexpr0.18\linewidth-2\tabcolsep\relax} >{\centering\arraybackslash}p{\dimexpr0.13\linewidth-2\tabcolsep\relax} >{\centering\arraybackslash}p{\dimexpr0.13\linewidth-2\tabcolsep\relax} >{\raggedright\arraybackslash}p{\dimexpr0.16\linewidth-2\tabcolsep\relax} >{\centering\arraybackslash}p{\dimexpr0.19\linewidth-2\tabcolsep\relax}}
\toprule
\textbf{Target} & \textbf{Sampled signals} & \textbf{COI evidence} & \textbf{Interfaces} & \textbf{Linked knobs} & \textbf{Reduction} \\
\midrule
AES-T1: ECB/AES-256/sideload & 3 sampled signals & 9 visited & 3 retained & 28 linked & 44 $\rightarrow$ 28 (36.4\%) \\
ES-T1: FIPS-scope cross & 6 sampled signals & 12 visited & 8 retained & 16 linked & 63 $\rightarrow$ 16 (74.6\%) \\
ES-T2: FIFO-state cross & 5 sampled signals & 39 visited & 27 retained & 41 linked & 63 $\rightarrow$ 41 (34.9\%) \\
\bottomrule
\end{tabular}
\end{table*}

\subsection{Worked AES Target}
\label{subsec:aes_case}
Table~\ref{tab:aes_case} traces one initially uncovered AES cross bin through the complete workflow.

\begin{table*}[!t]
\centering
\caption{End-to-end AES case study.}
\label{tab:aes_case}
\footnotesize
\renewcommand{\arraystretch}{1.12}
\begin{tabular}{>{\bfseries\raggedright\arraybackslash}p{\dimexpr0.21\linewidth-2\tabcolsep\relax} >{\raggedright\arraybackslash}p{\dimexpr0.79\linewidth-2\tabcolsep\relax}}
\toprule
\textbf{Case-study item} & \textbf{Audited content} \\
\midrule
Original bin & \nolinkurl{aes_cov_if::aes_ctrl_cg.cr_mode_key_len.ecb;aes_256;auto[1]}, the \texttt{cp\_mode} $\times$ \texttt{cp\_key\_len} $\times$ \texttt{cp\_sideload} cross requiring ECB, AES-256, and sideload $=1$; initial count 0. \\
DUT signals & \texttt{aes\_mode}, \texttt{aes\_keylen}, and \texttt{aes\_sideload}, linked to \texttt{reg2hw.ctrl\_shadowed.mode.q}, \texttt{key\_len.q}, and \texttt{sideload.q}. \\
COI path & Stored influential interfaces are \texttt{aes\_mode}, \texttt{aes\_keylen}, and \texttt{aes\_sideload}. Representative paths are mode $\rightarrow$ \texttt{ral.ctrl\_shadowed.mode.set} $\rightarrow$ \texttt{aes\_mode} $\rightarrow$ target and sideload $\rightarrow$ \texttt{ral.ctrl\_shadowed.sideload} $\rightarrow$ \texttt{aes\_sideload} $\rightarrow$ target, with promoted aliases leading to plusarg/configuration controls. \\
Linked knobs & Static linkage retained 28 target-linked knobs: 9 temporal and 19 constraint. These category counts enter the difficulty classifier. \\
Formal property & \texttt{c\_lmc\_target} uses \texttt{@(posedge clk\_i) disable iff (!rst\_ni)} with predicate \texttt{aes\_mode inside \{AES\_ECB\}} $\land$ \texttt{aes\_keylen inside \{AES\_256\}} $\land$ \texttt{aes\_sideload == 1'b1}, bound to the AES top level. \\
Triage result & Reachable in 2 cycles, with witness depth 2, proof depth 5, solver time 0.48 s, and full Jasper invocation 10.957 s. The classifier inputs $d=2$, $e=0$, $M=0$, $K_t=9$, $P=1$, and $K_c=19$ yield $s_T=9$ and $s_C=21$, so the target is \textsc{ConstraintHeavy}. It remains in the denominator and is routed to low-beta UCB ($\beta=0.35$). \\
Group-specific BO execution & Fifth executed candidate, \texttt{aes\_0004} (zero-based iteration 4, seed 5): ECB, AES-256, sideload enabled, \texttt{aes\_stress\_all}, \texttt{aes\_stress\_all\_vseq}, message length 1, reseed rate 64, manual-operation 100\%, config-error 0\%, error type 2. These 10 assigned executable knobs expand to 48 encoded features. The raw combinational pool had 40,551 candidates; constraint and incompatibility filters removed 4,175 and 36,255, leaving 121 to score. The configured 1,024 value is a per-restart sampling cap and was inactive here. Predicted mean 0.083121, uncertainty 0.521828, two transfer samples, and zero target-specific samples. \\
BO objective & $H=1$; stepping values $\{0,1,0,2\}$ give $S=0.75$; $F=1/(1+2/20)=0.909091$; $C=148.056561$ s; valid candidate. Hence $J=1+0.3(0.75)+0.2(0.909091)-0.001(148.056561)=1.258762$. \\
Final outcome & Simulation passed; coverage feedback was \texttt{direct\_hit} with target raw count 2. The bin is simulator-covered and belongs to post-cleanup group-specific BO. It is not formal-cleaned because the formal outcome was reachable. \\
\bottomrule
\end{tabular}
\end{table*}

\subsection{Objective-Weight Sensitivity}
\label{subsec:sensitivity}
The fixed implementation weights were not tuned per target. Table~\ref{tab:sensitivity} compares four matched 20-simulation settings over all retained targets. Only the objective weights differ.

\begin{table*}[!t]
\centering
\caption{Matched objective-weight sensitivity study: 20 simulations per DUT, 28 retained AES targets, and 150 retained Entropy Source targets. Hits are unique newly covered bins, and one simulation may cover multiple bins.}
\label{tab:sensitivity}
\small
\begin{tabular}{>{\raggedright\arraybackslash}p{\dimexpr0.13\linewidth-2\tabcolsep\relax} >{\centering\arraybackslash}p{\dimexpr0.09\linewidth-2\tabcolsep\relax} >{\centering\arraybackslash}p{\dimexpr0.09\linewidth-2\tabcolsep\relax} >{\centering\arraybackslash}p{\dimexpr0.09\linewidth-2\tabcolsep\relax} >{\centering\arraybackslash}p{\dimexpr0.12\linewidth-2\tabcolsep\relax} >{\raggedleft\arraybackslash}p{\dimexpr0.12\linewidth-2\tabcolsep\relax} >{\raggedleft\arraybackslash}p{\dimexpr0.12\linewidth-2\tabcolsep\relax} >{\raggedright\arraybackslash}p{\dimexpr0.24\linewidth-2\tabcolsep\relax}}
\toprule
\textbf{Setting} & $\lambda_H$ & $\lambda_S$ & $\lambda_F$ & $\lambda_C$ & \textbf{AES hits} & \textbf{ES hits} & \textbf{Purpose} \\
\midrule
Hit-only & 1.0 & 0 & 0 & 0.001 & 5 & 27 & Value of dense $S/F$ shaping \\
Default & 1.0 & 0.3 & 0.2 & 0.001 & 20 & 93 & Measured control \\
High-S & 1.0 & 0.6 & 0.2 & 0.001 & 13 & 55 & Stronger stepping emphasis \\
No-F & 1.0 & 0.3 & 0 & 0.001 & 21 & 74 & Value of formal alignment \\
\bottomrule
\end{tabular}
\end{table*}

\rTwoThree{Default produced more hits than Hit-only and High-S in these audited runs for both DUTs. Raising $\lambda_S$ from 0.3 to 0.6 reduced hits in these audited runs, consistent with excessive proximity emphasis distracting from confirmed closure in this configuration. Removing $F$ changed AES marginally (21 versus 20) but reduced Entropy Source from 93 to 74 hits; the observed effect of formal alignment differed between the two audited DUT runs. All four settings retain formal grouping and group-specific policy selection; No-F removes only the formal-alignment term from the objective and is not a triage-free comparison. Because each row is one matched fixed-budget run, these results are audited-run observations rather than replicated statistical estimates.}

\subsection{Runtime Interpretation}
\label{subsec:runtime_results}
The formal, AST/COI, BO model-fit, simulation, and total BO-stage timings in Table~\ref{tab:cost_scale} make the resource tradeoff visible. The measured 20-candidate BO-stage wall-clock\rTwoThree{, including simulation execution, is reported as total BO-stage wall-clock}, while mean simulation runtime is reported separately for context. Because these measurements come from one cluster configuration and license allocation, we do not claim that the timing transfers unchanged to another machine, simulator release, or license configuration.

\subsection{\rTwoOne{Endpoint Visualization Protocol}}
\label{subsec:trend_plot_setup}
\label{subsec:illustrative_trend_plots}
\rTwoOne{Figures~\ref{fig:aes_endpoints} and~\ref{fig:es_endpoints} use an endpoint-only bar-chart protocol. Panel (a) compares the audited raw-coverage endpoints of the matched search branches. Panel (b) reports sequential coverage accounting before and after complete-proof cleanup and post-cleanup BO: solid blue bars use the fixed original denominator, while orange hatched bars use the same simulator-confirmed numerator with the effective denominator. Cleanup changes the AES denominator from 271 to 269 and the Entropy Source denominator from 2277 to 2267 without changing either numerator. No intermediate point, connecting line, or inferred trajectory is shown, so the figures do not assert convergence evidence between endpoints. The audited hit totals and endpoint values remain those in Tables~\ref{tab:matched_search} and~\ref{tab:sequential_accounting}.}

\begin{figure*}[!t]
\centering
\includegraphics[width=\textwidth]{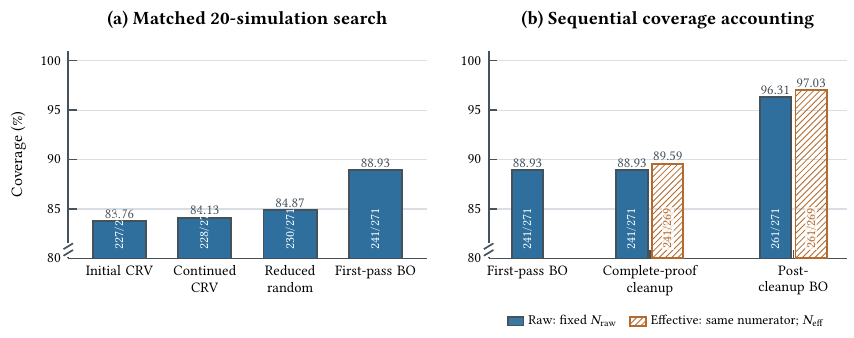}
\Description{AES endpoint coverage comparison using bar charts. Panel (a) reports raw functional-coverage endpoints for the initial CRV, continued CRV, reduced-random, and first-pass BO branches under the matched 20-simulation search budget. Panel (b) reports raw and effective coverage endpoints for first-pass BO, complete-proof cleanup, and post-cleanup BO. Solid blue bars use the fixed original denominator, orange hatched bars use the same numerator with the effective denominator, and cleanup changes the denominator from 271 to 269 without changing the numerator.}
\caption{\rTwoOne{AES endpoint coverage comparison. Panel (a) reports audited raw-coverage endpoints for the matched 20-simulation search branches. Panel (b) reports sequential raw/effective coverage accounting for first-pass BO, complete-proof cleanup, and post-cleanup BO. Solid blue bars use the fixed original denominator, while orange hatched bars use the same numerator with the effective denominator; cleanup changes only the denominator from 271 to 269 and adds no raw hits. No intermediate trajectory is shown.}}
\label{fig:aes_endpoints}
\end{figure*}

\begin{figure*}[!t]
\centering
\includegraphics[width=\textwidth]{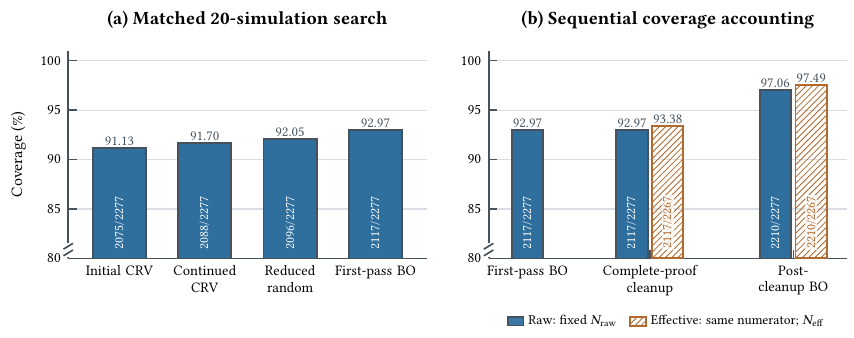}
\Description{Entropy Source endpoint coverage comparison using bar charts. Panel (a) reports raw functional-coverage endpoints for the initial CRV, continued CRV, reduced-random, and first-pass BO branches under the matched 20-simulation search budget. Panel (b) reports raw and effective coverage endpoints for first-pass BO, complete-proof cleanup, and post-cleanup BO. Solid blue bars use the fixed original denominator, orange hatched bars use the same numerator with the effective denominator, and cleanup changes the denominator from 2277 to 2267 without changing the numerator.}
\caption{\rTwoOne{Entropy Source endpoint coverage comparison. Panel (a) reports audited raw-coverage endpoints for the matched 20-simulation search branches. Panel (b) reports sequential raw/effective coverage accounting for first-pass BO, complete-proof cleanup, and post-cleanup BO. Solid blue bars use the fixed original denominator, while orange hatched bars use the same numerator with the effective denominator; cleanup changes only the denominator from 2277 to 2267 and adds no raw hits. No intermediate trajectory is shown.}}
\label{fig:es_endpoints}
\end{figure*}
\end{revisionadd}

\section{Discussion and Limitations}
\label{sec:discussion}
\begin{revisionadd}
\paragraph{Attribution of gains.}
The matched baselines \rTwoThree{provide a bounded observation of DUT-dependent BO behavior: in these audited fixed-budget runs, BO produced} 11 AES and 21 Entropy Source hits relative to random search over the same reduced space. These search-only baselines are matched by simulation count rather than wall-clock time. Formal cleanup contributes no numerator hits and only refines the effective denominator. The later 20 and 93 hits are joint outcomes of grouping, group-policy selection, BO proposal, and simulation. Because no matched \rTwoOne{default-BO control after the same formal cleanup was run}, the post-cleanup gains cannot be attributed causally to \rTwoOne{classification or routing alone}. Thus the evidence supports the combined diagnosis-and-search workflow, while classification and denominator cleanup themselves create no coverage.

\paragraph{Coverage-model dependence.}
The workflow optimizes an existing functional coverage model. Closing that model does not establish that the coverage plan is complete or aligned with every requirement in the design specification. Missing or poorly designed coverpoints remain outside the method's scope, and the final percentages must not be interpreted as specification-level verification completeness.

\paragraph{Formal assumptions, automation, and scalability.}
Every reachability conclusion is conditional on the evaluated RTL and the stated environment assumptions. The over-constraint audit reduces, but cannot eliminate, the risk of an assumption mismatch. Bounded, timed-out, unknown, and error outcomes remain in the denominator. The evaluated formal problems are target-specific IP/block-level queries; full-chip temporal targets may require substantially greater resources. The formal stage is semi-automatic because proof completeness and assumption adequacy remain audit-reviewed.

\paragraph{Training history and objective sensitivity.}
BO has no separate offline training phase. Warm-start information comes from summaries of the 3,000-run and 10,000-run CRV histories, and candidate-specific observations accumulate online. These corpus sizes are not the number of surrogate-training samples. The four weight settings are matched fixed-budget runs without statistical replication, so their results establish implementation sensitivity only.

\paragraph{External validity and reproducibility.}
The evaluation covers two OpenTitan IPs, one cluster configuration, no controlled BO run over the full unreduced space, and no matched \rTwoOne{default-BO control after the same formal cleanup}. A direct numerical comparison with a published framework was not included because no prior framework offered a shared DUT, coverage model, and executable artifact for a controlled benchmark; continued CRV and reduced-space random search serve as common-reference controls instead. Broader multi-DUT and cross-platform studies remain future work. The full internal harness is not released at this stage. The workflow pseudocode, property-generation rules, schemas, parameters, assumptions, and audited case data support independent reimplementation, but not exact artifact-level reproduction; equivalent simulator and formal tooling remains necessary.
\end{revisionadd}

\section{Conclusion}
\label{sec:conclusion}
\begin{revisionadd}
This article presents a non-intrusive, staged workflow for closing residual bins in an existing functional coverage model. AST/COI analysis identifies the available control surface, matched first-pass BO searches the reduced knob space, complete formal proofs refine only the effective denominator, and formal difficulty labels route retained targets to group-specific BO. Simulator coverage artifacts remain the sole source of numerator hits.

Under the matched first-pass search budget, BO reaches 241/271 AES bins (88.93\% raw coverage) and 2117/2277 Entropy Source bins (92.97\% raw coverage). Complete-proof cleanup removes two and ten bins from the respective effective denominators without adding coverage. \rTwoThree{The integrated post-cleanup stage---difficulty grouping, policy routing, and group-specific BO---adds 20 and 93 simulator-confirmed hits.} Final raw coverage is therefore 261/271 (96.31\%) for AES and 2210/2277 (97.06\%) for Entropy Source; final effective coverage is 261/269 (97.03\%) and 2210/2267 (97.49\%).

The results support a bounded \rTwoThree{observation: in these audited fixed-budget runs, the diagnosis-and-search workflow improved} last-mile closure on the two evaluated OpenTitan IPs while preserving their mature UVM environments. They do not establish coverage-plan completeness, full-chip formal scalability, or cross-platform timing equivalence. Future work will extend the evaluation to additional DUTs, add replicated sensitivity trials, and establish a shared benchmark for controlled cross-framework comparison.
\end{revisionadd}

\bibliographystyle{ACM-Reference-Format}
\bibliography{todaes_bo_formal_front}

\end{document}